\documentclass[aps,pra,reprint,superscriptaddress,floatfix]{revtex4-1}
\usepackage{blindtext}
\usepackage{amsfonts}
\usepackage{amsmath}
\usepackage{amssymb}
\usepackage{graphicx}%
\usepackage{color}
\usepackage{epstopdf}

\DeclareSymbolFont{myletters}{OML}{ztmcm}{m}{it}
\DeclareMathSymbol{\uplambda}{\mathord}{myletters}{"15}

\begin{document}
\title{First-principles study of superconductivity in 2D and 3D forms of PbTiSe$_{2}$: Suppressed charge density wave in 1\emph{T}-TiSe$_{2}$}

\author{Bao-Tian Wang}
\thanks{Author to whom correspondence should be addressed. E-mail: wangbt@ihep.ac.cn }
\affiliation{Institute of High Energy Physics, Chinese Academy of
Sciences (CAS), Beijing 100049, China} \affiliation{Dongguan Neutron
Science Center, Dongguan 523803, China}
\author{Peng-Fei Liu}
\affiliation{Institute of High Energy Physics, Chinese Academy of
Sciences (CAS), Beijing 100049, China} \affiliation{Dongguan Neutron
Science Center, Dongguan 523803, China}
\author{Jing-Jing Zheng}
\affiliation{Institute of High Energy Physics, Chinese Academy of
Sciences (CAS), Beijing 100049, China} \affiliation{Dongguan Neutron
Science Center, Dongguan 523803, China}
\author{Wen Yin}
\affiliation{Institute of High Energy Physics, Chinese Academy of
Sciences (CAS), Beijing 100049, China} \affiliation{Dongguan Neutron
Science Center, Dongguan 523803, China}
\author{Fangwei Wang}
\affiliation{Dongguan Neutron Science Center, Dongguan 523803,
China} \affiliation{Beijing National Laboratory for Condensed Matter
Physics, Institute of Physics, Chinese Academy of Sciences, Beijing
100080, China}

\begin{abstract}
Layered 1$T$-TiSe$_{2}$ has attracted much interest for the competition of charge density wave (CDW) and superconductivity in its bulk and even monolayer forms. Here we perform first-principles calculations of the electronic structure, phonon dispersion, and electron-phonon coupling of the Pb-intercalated 1$T$-TiSe$_{2}$ in bulk and layered structures. Results show that upon the Pb atom intercalation, the CDW instability in 1$T$-TiSe$_{2}$ can be effectively suppressed, accompanied by the removal of the imaginary phonon modes at \textbf{q}$_{\rm{M}}$. The Pb 6\emph{p} orbitals occupy directly at the Fermi level, which hence intercalates the superconductivity. Both bulk and layered PbTiSe$_{2}$ are phonon-mediated superconductors, with estimated superconducting temperature $T_{c}$ to be $\sim$1.6-3.8 K. The main contribution to the electron-phonon coupling is from the vibrations of Pb and Se atoms. The superconducting related physical quantities are found tunable by varying Pb content.

\end{abstract}

\maketitle

\section{INTRODUCTION}
Since the successful fabrication of the monolayer of the transition metal dichalcogenides (TMDs) using a micromechanical cleavage method \cite{novoselov2005two}, layered TMDs consisting of two-dimensional (2D) sheets bonded through weak van der Waals (vdW) forces
have attracted much attention \cite{lebegue2009electronic,xiao2012coupled,shi2013quasiparticle,paul2017computational,mounet2018two} owing to their potential applications
in electronics \cite{fiori2014electronics}, photonics \cite{mak2010atomically}, sensor \cite{perea2014cvd}, energy storages \cite{da2014supercapacitor}, and catalysis \cite{ho2004preparation} as well as their fascinating physical properties of superconductivity \cite{shi2015superconductivity}, CDW \cite{klemm2015pristine}, and topological surface/edge states \cite{bahramy2018ubiquitous,liu2016new}. Among these TMDs, several of them show competition or coexistence of the CDW and superconducting order in three-dimensional (3D) bulk and/or even the monolayer limit, such as 2\emph{H}-NbSe$_{2}$ \cite{weber2011extended}, 1$T$-TaS$_{2}$ \cite{sipos2008mott}, IrTe$_{2}$ \cite{cao2013origin}, ZrTe$_{3}$ \cite{hoesch2016evolution}, and 1$T$-TiSe$_{2}$ \cite{morosan2006superconductivity}. The interplay between these two fundamental electronic phenomena, existing strong debates in condensed matter physics, requires being further understood.

Both bulk and monolayer 1$T$-TiSe$_{2}$ undergo CDW instability at around 205-230 K characterized by 2$\times$2$\times$2 or 2$\times$2$\times$1 superlattices \cite{chen2015charge,holt2001x}. Although pristine TiSe$_{2}$ is not superconducting at low temperature, suppression of its CDW and stabilization of the superconductivity can be realized by Cu intercalation \cite{morosan2006superconductivity}, pressure \cite{kusmartseva2009pressure}, or electrostatic gating \cite{li2016controlling}. In some recent experiments \cite{kogar2017observation,yan2017influence}, it was reported that an incommensurate CDW phase may be an important precursor to the superconducting dome in Cu$_{x}$TiSe$_{2}$ (\emph{x}$\sim$0.06-0.09). Thus, whether the CDW and the superconducting order in 1$T$-TiSe$_{2}$ exhibit as competition or coexistence is still in exploring. Based upon the Fermi surface and band structure results from the angle resolved photoemission spectroscopy (ARPES) \cite{zhao2007evolution} as well as the density functional theory (DFT) \cite{calandra2011charge}, the Ti 3\emph{d} band was believed playing crucial role in the superconductivity in both Cu-intercalated TiSe$_{2}$ \cite{li2007single} and compressed pristine TiSe$_{2}$ \cite{kusmartseva2009pressure}. Calandra \emph{et al.} \cite{calandra2011charge} stated that a stiffening of the short-range force constants, other than the nesting effects, is responsible for the disappearing of the CDW under pressure.

In the present work, we perform first-principles calculations of the electronic structures, lattice dynamics, and electron-phonon coupling (EPC) of the bulk and 2D forms for Pb-intercalated 1$T$-TiSe$_{2}$, so as to investigate the effects of heavy-metal doping on the CDW and superconductivity. The doping content of Pb in our work is greatly larger than that of Cu in Cu$_{x}$TiSe$_{2}$, where the Cu composition \emph{x} only varies in range of 0-0.1 \cite{morosan2006superconductivity,kogar2017observation,yan2017influence}. Our calculated results indicate that the CDW instability in 2D and 3D forms can be suppressed effectively by the Pb intercalation. A visible doping of the Pb 6\emph{p} electrons is responsible for this CDW suppression and the strong EPC contribution from Pb vibrations plays key role for the superconductivity. Increasing the Pb content from a composition of 0.5 to 1, the EPC and the superconducting transition temperature $T_{c}$ of the Pb$_{x}$TiSe$_{2}$ monolayer are greatly enhanced.
\section{COMPUTATIONAL DETAILS}
The calculations are performed at the DFT level, employing the Perdew-Burke-Ernzerhof (PBE) and projector augmented-wave (PAW) pseudopotentials
\cite{blochl1994projector} as implemented in the QUANTUM-ESPRESSO package \cite{giannozzi2009quantum}. The non-local vdW
density functional optB86b-vdW \cite{klimevs2009chemical,klimevs2011van},
which has long been recognized very important in predicting ground states
and describing the interlayer distance in many classes of
materials \cite{sa2016development,wang2017evolution,mcguire2017antiferromagnetism,wu2017engineering}, is utilized to properly
treat the long-range dispersive interactions. The plane-waves kinetic-energy cutoff is
set as 50 Ry and the structural optimization is performed until the
forces on atoms are less than 10 meV/\AA. Monolayer 1$T$-TiSe$_{2}$, Pb-intercalated bilayer TiSe$_{2}$ (TSPTS), and Pb-decorated monolayer TiSe$_{2}$ (PTS and PTSP) are simulated
with a vacuum thickness of 20 \AA, which is enough to decouple the
adjacent layers. Here, the PTS refers to one Pb atom decorated on one side of monolayer TiSe$_{2}$ while the PTSP refers to two Pb atoms decorated on both sides of the monolayer. Unshifted Brillouin-zone (BZ) \textbf{k}-point
meshes of 18$\times$18$\times$8 and 18$\times$18 are
adopted for the electronic charge density calculations for 3D bulk and 2D monolayer, respectively. The phonon
modes are computed within density-functional perturbation theory
\cite{baroni2001phonons} on 6$\times$6$\times$4 and 6$\times$6 \textbf{q} meshes for 3D bulk and 2D monolayer, respectively. The relativistic effects in terms of the spin-orbit coupling (SOC) are included self-consistently in the electronic structure calculations. In calculating the phonon dispersions, since the effect of SOC is less important in describing the vibrational properties \cite{wang2013phonon,zheng2017electron,wei2017manipulating,hou2017structural}, we neglect this effect.

\section{RESULTS AND DISCUSSIONS}

\begin{figure}[ptb]
	\centering
	\includegraphics[width=1\linewidth]{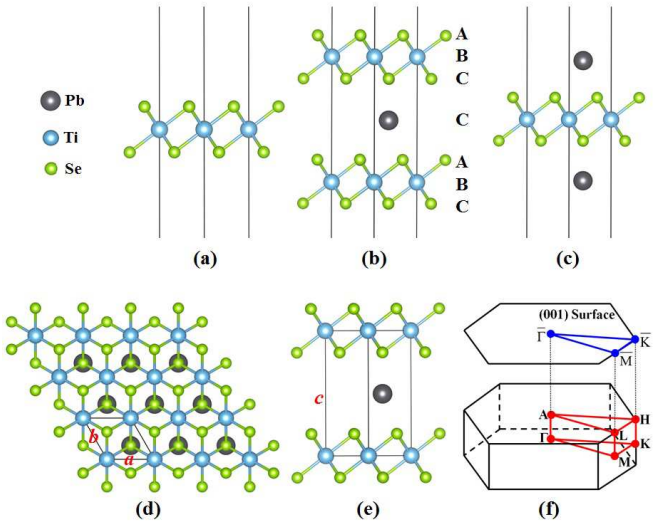}
	\caption{Side view of (a) monolayer 1$T$-TiSe$_{2}$, (b) Pb-intercalated bilayer TiSe$_{2}$ (TSPTS), and (c) Pb-decorated monolayer TiSe$_{2}$ (PTSP). (d),(e) Top and side view of bulk PbTiSe$_{2}$. (f) Bulk and (001) surface BZ of TiSe$_{2}$ and PbTiSe$_{2}$.}
	\label{fgr:fig1}
\end{figure}

\begin{figure}[ptb]
\begin{center}
\includegraphics[width=1\linewidth]{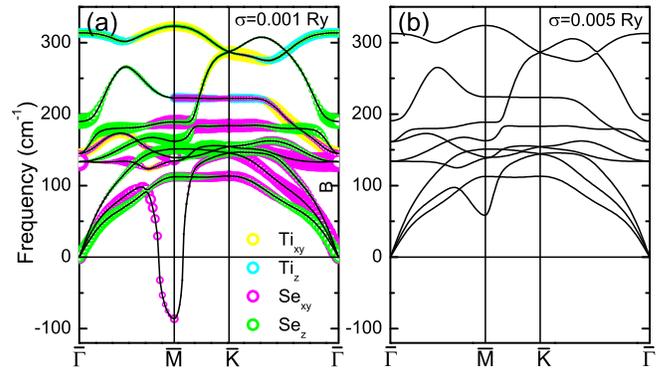}
\end{center}
\caption{Phonon dispersions of monolayer 1$T$-TiSe$_{2}$ calculated by using Fermi-Dirac smearing widths of (a) 0.001 Ry and (b) 0.005 Ry. The yellow, cyan, magenta, and green hollow circles in (a) indicate Ti
horizontal, Ti vertical, Se horizontal, and Se vertical modes, respectively.}%
\label{fig3}%
\end{figure}

\begin{figure}[ptb]
	\centering
	\includegraphics[width=9cm]{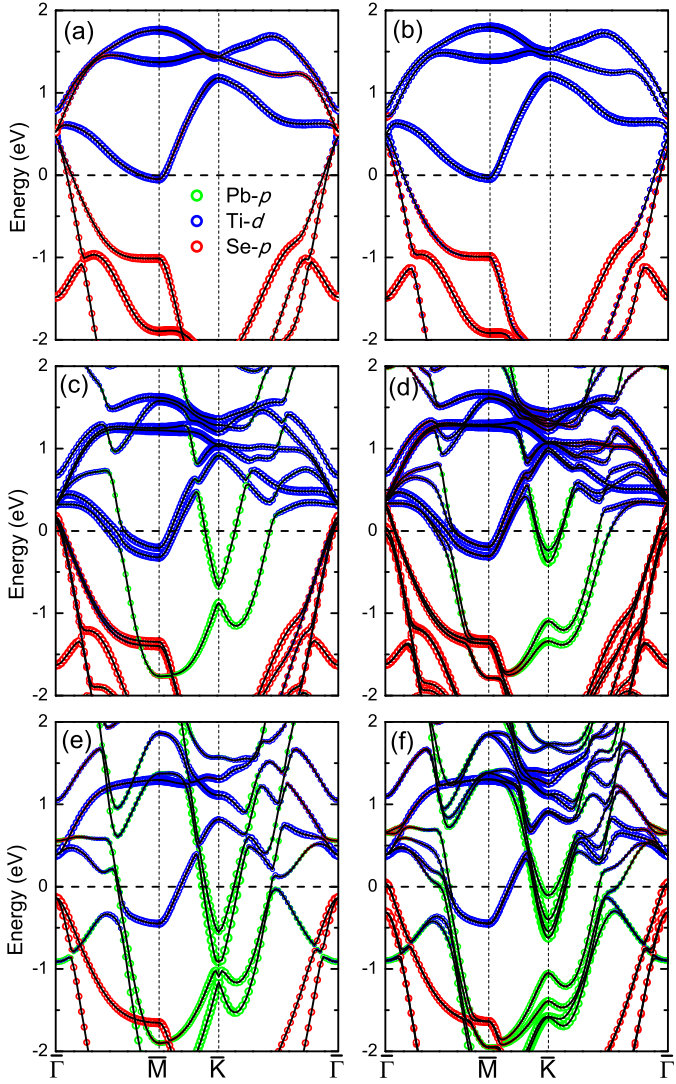}
	\caption{Orbital-resolved band structures of (a)(b) monolayer 1$T$-TiSe$_{2}$, (c)(d) TSPTS, and (e)(f) PTSP calculated without SOC (left one) and with SOC (right one).}
	\label{fgr:fig2}
\end{figure}

\begin{figure*}[ptb]
	\centering
	\includegraphics[width=0.9\linewidth]{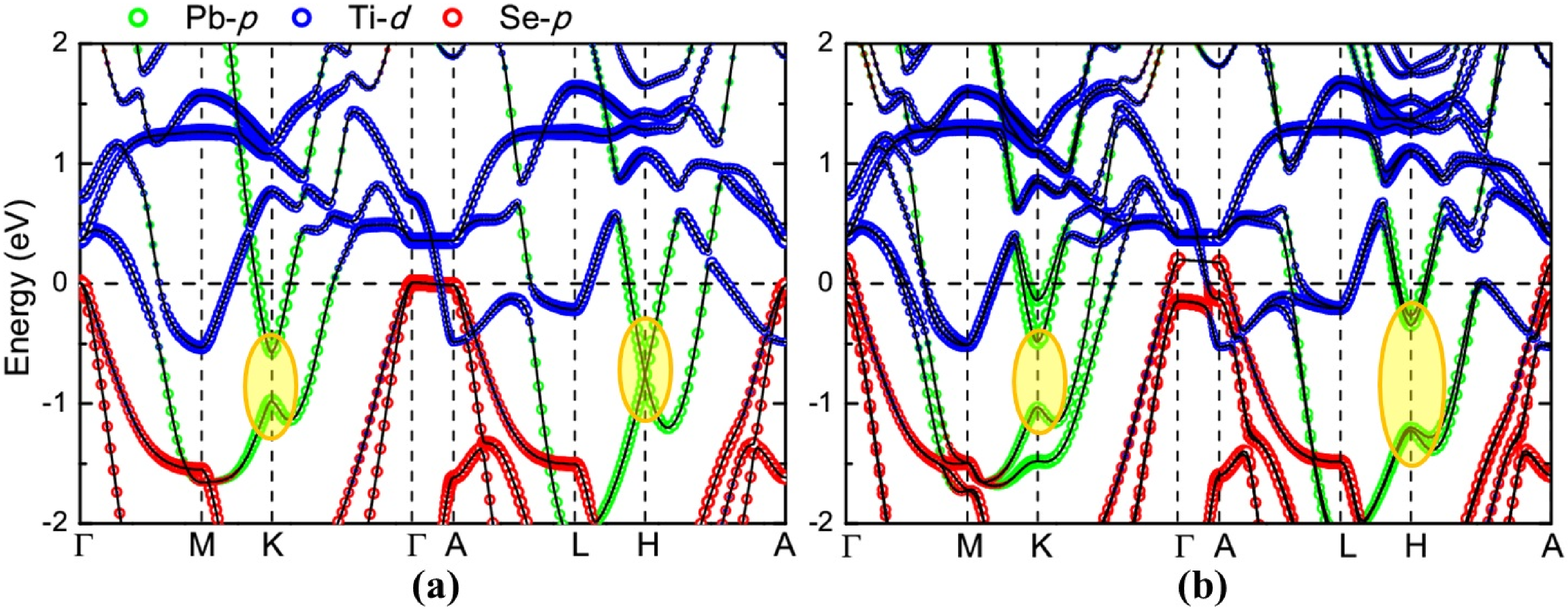}
	\caption{Orbital-resolved band structures of bulk PbTiSe$_{2}$ calculated (a) without SOC and (b) with SOC.}
	\label{fgr:fig3}
\end{figure*}
We present in Figs. 1(a)-1(f) the crystal structures of monolayer 1$T$-TiSe$_{2}$, TSPTS, PTSP, and bulk PbTiSe$_{2}$ along with the corresponding BZ. In each monolayer 1$T$-TiSe$_{2}$, the Ti atoms are surrounded by the nearest six Se atoms to form an octahedron with the triple layers (TLs) in ABC stacking order. In bulk 1$T$-TiSe$_{2}$, the atomic structure can be visualized as a superposition of the TLs along the
$c$-axis in ABC-ABC stacking. Owing to the weak coupling between two adjacent TLs, it has been successful to
achieve monolayer 1$T$-TiSe$_{2}$ by exfoliating the bulk 1$T$-TiSe$_{2}$ \cite{goli2012charge} or
using molecular beam epitaxy \cite{chen2015charge,peng2015molecular}. In obtaining correct structure for bulk PbTiSe$_{2}$, we test two possible stacking orders: one is ABC-C-ABC and another ABC-B-ABC. We find that the former one is energetically (by 69 meV per formula unit) more stable than the latter one. Thus, in the following we only consider the ABC-C-ABC stacking in both 2D and 3D structures of PbTiSe$_{2}$ (See Fig. 1). The optimized lattice constants of the bulk PbTiSe$_{2}$ are $a$ = 3.514 \AA{} and $c$ = 8.938 \AA{} and the bond lengths of Ti-Se and Pb-Se are 2.556 and 2.962 \AA{}, respectively. The value of our optimized in-plane lattice constant $a$ accords well with recent PAW-PBE result of 3.51 \AA{} \cite{fu2016modulation} and experimental value of 3.540 \AA{} \cite{riekel1976structure} for bulk 1$T$-TiSe$_{2}$.

Upon cooling, the monolayer and bulk 1$T$-TiSe$_{2}$ respectively undergo
a second-order CDW transition at around 230 and 205 K to the commensurate 2$\times$2$\times$1 and 2$\times$2$\times$2 superlattices \cite{chen2015charge,holt2001x}. Below the CDW transition temperature, a negative phonon mode at vector \textbf{q}$_{\rm{M}}$ or \textbf{q}$_{\overline{\rm{M}}}$ has been reported for the high-symmetry structures of the monolayer and bulk 1$T$-TiSe$_{2}$ in some recent DFT works \cite{fu2016modulation,wei2017manipulating,singh2017stable}. In our present work, as shown in Fig. 2(a), we also observe this kind of negative acoustic mode with a Fermi-Dirac smearing width of $\sigma$ = 0.001 Ry. From the decomposition of the phonon spectrum with respect to Ti and Se atomic vibrations, we find that the instability at \textbf{q}$_{\overline{\rm{M}}}$ is mainly induced by the in-plane Se$_{xy}$ vibrations. The main contribution to the phonon modes below 200 cm$^{-1}$ is from Se vibrations while above that is from Ti atoms. By increasing the electronic temperature, changing the Fermi-Dirac smearing width $\sigma$ to 0.005 Ry, the negative phonon at \textbf{q}$_{\overline{\rm{M}}}$ is manipulated to positive [Fig. 2(b)], but still exhibiting soften frequencies.

Suppressing the CDW state in 1$T$-TiSe$_{2}$ by Cu intercalation \cite{morosan2006superconductivity}, pressure \cite{kusmartseva2009pressure}, and electron/hole doping \cite{wei2017manipulating} can result in a superconducting state. To understand how superconductivity can compete with the CDW instability in PbTiSe$_{2}$, in the following, we will investigate the electronic structure, vibration properties and the EPC of Pb-intercalated TiSe$_{2}$ in 2D and 3D forms.

We present in Fig. 3 the orbital-resolved band structures of monolayer 1$T$-TiSe$_{2}$, TSPTS, and PTSP and in Fig. 4 the band structures of bulk PbTiSe$_{2}$. We compare results obtained with and without SOC. For monolayer 1$T$-TiSe$_{2}$, the Ti 3$d$ and Se 4$p$ orbitals dominate the bands around the Fermi energy level. The hole pocket at the $\overline{\Gamma}$ point is mainly featured by the Se 4$p$ whereas the electron pocket at the M point is characterized by the Ti 3$d$. The SOC effect on these bands is invisible. After intercalation or decoration of Pb atoms, as shown in Figs. 3(c)-3(f), both the hole pocket at the $\overline{\Gamma}$ point and the electron pocket at the $\overline{\rm{M}}$ point are shifted downwards. A new hole pocket dominated by Pb 6$p$ orbitals appears at the $\overline{\rm{K}}$ point. The SOC effect on the Pb-dominated bands is visible. For example, the Dirac-like band gap at the $\overline{\rm{K}}$ point at around -0.8 eV for TSPTS is greatly enlarged by the SOC from 0.208 to 0.716 eV. As indicated by the elliptical yellow shades in Fig. 4, the Dirac-like band gap at the $\rm{K}$ point is enlarged and the Dirac point at the $\rm{H}$ point is opened by inclusion of the SOC. The Dirac point here may induce appearance of the surface state (SS) \cite{wang2017evolution,chang2016topological} in PbTiSe$_{2}$ slabs. Upon increasing the content of Pb from TSPTS to PTSP, more and more Pb-dominated bands appear and cross the Fermi energy level, especially near the $\overline{\rm{K}}$ point.

\begin{figure*}[ptb]
	\centering
	\includegraphics[width=0.75\linewidth]{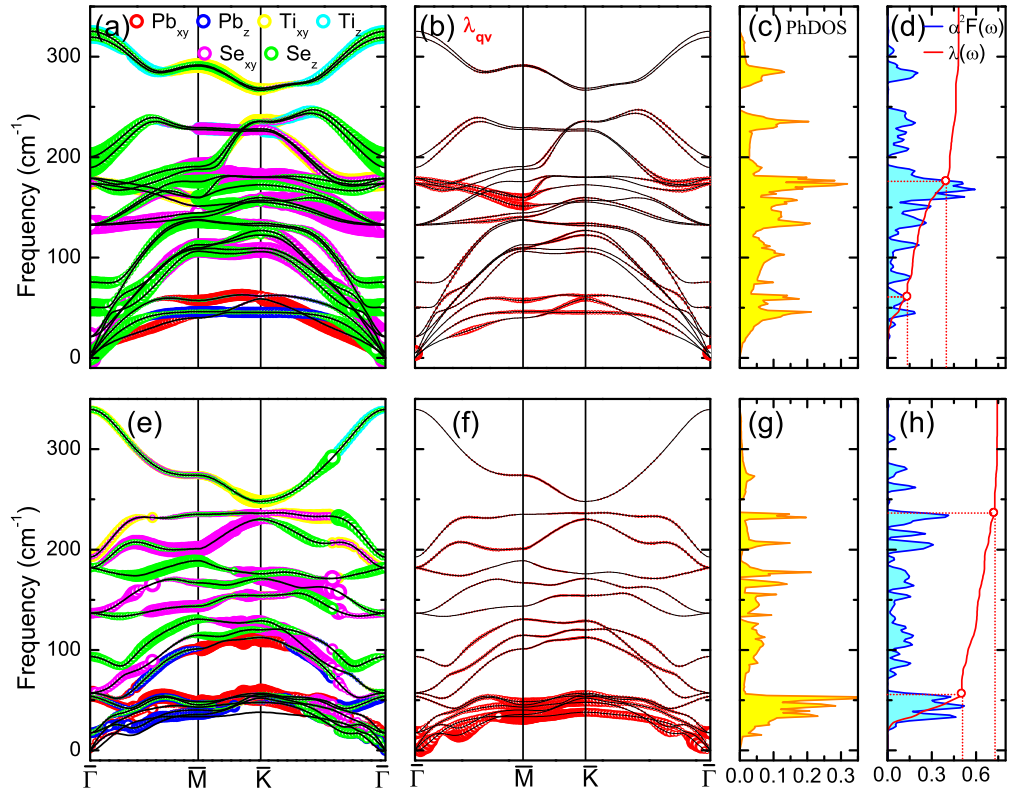}
	\caption{The phonon dispersions, PhDOS, Eliashberg spectral function $\alpha^{2}$F($\omega$), and cumulative
frequency-dependent of EPC $\uplambda$($\omega$) of (a)-(d) TSPTS and (e)-(h) PTSP. The phonon dispersions are weighted
by the motion modes of Pb, Ti, and Se atoms as well as the magnitude
of the EPC $\uplambda_{\textbf{\emph{q}}\nu}$ in the first-left and the second-left panels, respectively. The red, blue, yellow, cyan, magenta, and green hollow circles in (a) and (e) indicate Pb horizontal, Pb vertical, Ti
horizontal, Ti vertical, Se horizontal, and Se vertical modes, respectively. The magnitude of $\uplambda_{\textbf{\emph{q}}\nu}$ is displayed with identical scale in all figures for comparison.}
	\label{fgr:fig3}
\end{figure*}

\begin{figure*}[ptb]
	\centering
	\includegraphics[width=1\linewidth]{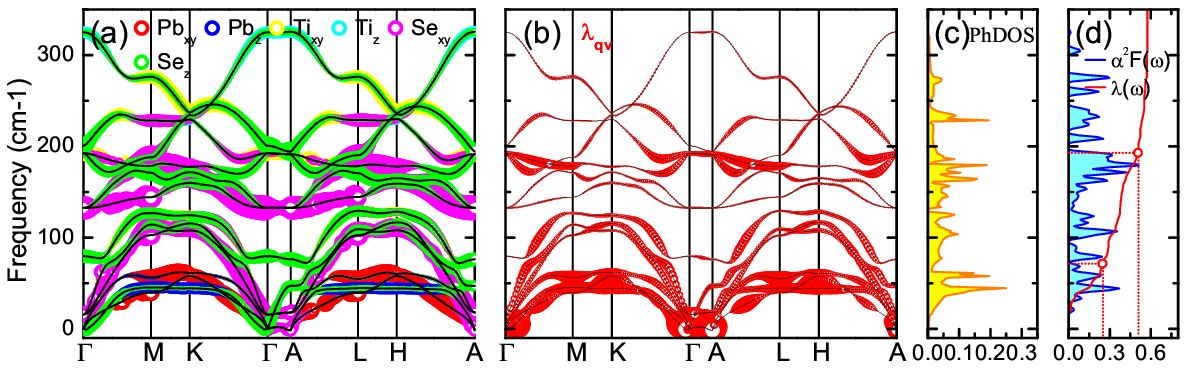}
	\caption{The phonon dispersions, PhDOS, Eliashberg spectral function $\alpha^{2}$F($\omega$), and cumulative
frequency-dependent of EPC $\uplambda$($\omega$) of bulk PbTiSe$_{2}$. The phonon dispersions are weighted
by the motion modes of Pb, Ti, and Se atoms as well as the magnitude
of the EPC $\uplambda_{\textbf{\emph{q}}\nu}$ in (a) and (b), respectively. The red, blue, yellow, cyan, magenta, and green hollow circles in (a) indicate Pb horizontal, Pb vertical, Ti
horizontal, Ti vertical, Se horizontal, and Se vertical modes, respectively. The magnitude of $\uplambda_{\textbf{\emph{q}}\nu}$ is displayed with an identical scale as in figure 5 for comparison.}
	\label{fgr:fig3}
\end{figure*}

We now focus on the vibration properties and the EPC in 2D and 3D forms of PbTiSe$_{2}$. Figures 5(a) and 5(e) show the phonon dispersions over the whole BZ for TSPTS and PTSP, respectively. The absence of the imaginary modes
clearly indicates that the 2D monolayer of PbTiSe$_{2}$ is dynamically stable, no matter in the case of low content (TSPTS) or in the case of high content (PTSP) of Pb. We confirm this fact by using very small value of the smearing width of $\sigma$ = 0.001 Ry. The doping of metal atoms (like Cu \cite{morosan2006superconductivity}) or electrons \cite{wei2017manipulating} can easily stabilize the CDW order in 1$T$-TiSe$_{2}$. Thus, our prediction here is reliable and also realizable in experiments in the future. In the case of low content of Pb, we find that the main contribution to the acoustic branches below 60 cm$^{-1}$ is from Pb vibrations. The in-plane vibrations of Se dominate in the
intermediate-frequency region from 60 to 250 cm$^{-1}$ while the out-of-plane Se$_{z}$
vibrations spread over the full BZ. The phonon modes of Ti atoms occupy the high frequencies above 270
cm$^{-1}$. In the case of high content of Pb, the main contribution to the acoustic branches below 60 cm$^{-1}$ is also from Pb. The Pb vibrations also occupy to some extent the first transverse optical
and longitudinal optical branches. The vibrations of Se are visible in wide area of the whole BZ while that of Ti only dominate the highest optical mode.

The phonon dispersions weighted by the magnitude of the EPC $\uplambda_{\textbf{\emph{q}}\nu}$, the phonon density of state (PhDOS), the Eliashberg electron-phonon spectral function $\alpha^{2}$F($\omega$), and the cumulative
frequency-dependent of EPC $\uplambda$($\omega$) are displayed in Figs. 5(b)-5(d) and 5(f)-5(h) for TSPTS and PTSP, respectively. Here, the $\uplambda_{\textbf{\emph{q}}\nu}$ is calculated, according to the Migdal-Eliashberg theory \cite{grimvall1981electron,giustino2017electron}, by $\uplambda_{\textbf{\emph{q}}\nu}$=$\frac{\gamma_{\textbf{\emph{q}}\nu}}{\pi\!hN(E_{\mathrm{{F}}})\omega_{\textbf{\emph{q}}\nu}^{2}}$, where $\gamma_{\textbf{\emph{q}}\nu}$ is the phonon linewidth, $\omega_{\textbf{\emph{q}}\nu}$ is the phonon frequency, and $N$(\emph{E}$_{\mathrm{{F}}}$) is the electronic density of state at the Fermi level. The $\gamma_{\textbf{\emph{q}}\nu}$ can be estimated by
\begin{align}
\gamma_{\textbf{\emph{q}}\nu}=\frac{2\pi\omega_{\textbf{\emph{q}}\nu}}{\Omega_{\rm{BZ}}}\sum_{\textbf{\emph{k}},n,m}|\rm{g}_{\textbf{\emph{k}}n,\textbf{\emph{k}}+\textbf{\emph{q}}m}^{\nu}|^{2}\delta(\epsilon_{\textbf{\emph{k}}n}-\epsilon_{F})\delta(\epsilon_{\textbf{\emph{k}}+\textbf{\emph{q}}m}-\epsilon_{F}),
\end{align}
where $\Omega_{\rm{BZ}}$ is the volume of BZ, $\epsilon_{\textbf{\emph{k}}n}$ and $\epsilon_{\textbf{\emph{k}}+\textbf{\emph{q}}m}$ denote the Kohn-Sham energy, and $\rm{g}_{\textbf{\emph{k}}n,\textbf{\emph{k}}+\textbf{\emph{q}}m}^{\nu}$ represents the EPC matrix element. The $\rm{g}_{\textbf{\emph{k}}n,\textbf{\emph{k}}+\textbf{\emph{q}}m}^{\nu}$, which can be determined self-consistently by the linear response theory, describes the probability amplitude for the scattering of an electron with a transfer of crystal momentum $\textbf{\emph{q}}$ \cite{allen1975transition}. The $\alpha ^{2}F(\omega)$ and the
$\uplambda$($\omega$) can be determined by
\begin{align}
\alpha^{2}F(\omega)=\frac{1}{2\pi\!N(E_{\mathrm{{F}}})}\sum_{\textbf{\emph{q}}\nu}\frac
{\gamma_{\textbf{\emph{q}}\nu}}{\omega_{\textbf{\emph{q}}\nu}}\delta(\omega-\omega_{\textbf{\emph{q}}\nu})
\end{align}
and
\begin{align}
\uplambda(\omega)=2\int_{0}^{\omega}\frac{\alpha^{2}\emph{F}(\omega)}{\omega}d\omega,
\end{align}
respectively. We find that in TSPTS the low-frequency phonons, mainly
associated with the Pb vibrations, account for 0.13 (27\%) of the
total EPC ($\uplambda$=0.48). The phonons in the intermediate-frequency region, dominated by Se modes, contributes an EPC strength of 0.27 (56\%). The large EPC strength here is mainly originated from the large values of the $\uplambda_{\textbf{\emph{q}}\nu}$ along the $\overline{\Gamma}$-$\overline{\rm{M}}$ direction in frequency range of 150-180 cm$^{-1}$, which has resulted in the largest peak of the PhDOS and $\alpha^{2}$F($\omega$). Similar to the Na-intercalated bilayer NbSe$_{2}$ \cite{lian2017first}, the EPC
induced by high-frequency phonons is almost negligible. As for PTSP, the phonons below 55 cm$^{-1}$ contribute 0.50 (68\%) of the total EPC ($\uplambda$=0.74). The phonons in frequency range of 55-236 cm$^{-1}$ only contribute 30\%. Overall, increasing the Pb content clealy enhance the EPC strength and our calculated
EPC value of 0.48 (0.74) makes the TSPTS (PTSP) a weak conventional superconductor.

In the case of the 3D form, as indicated by the phonon dispersion, PhDOS, $\alpha^{2}$F($\omega$), and $\uplambda$($\omega$) of bulk PbTiSe$_{2}$ in Fig. 6, the Pb vibrations dominate the low-frequency region below 72 cm$^{-1}$ with large EPC $\uplambda_{\textbf{\emph{q}}\nu}$ and account for 0.24 (42\%) of the
total EPC ($\uplambda$=0.57). The Se vibrations contribute mainly the intermediate-frequency region (72-194 cm$^{-1}$) with moderate magnitudes of the EPC $\uplambda_{\textbf{\emph{q}}\nu}$ and account for an EPC strength of 0.27 (47\%). The EPC value of 0.57, in between that of the TSPTS and PTSP, makes bulk PbTiSe$_{2}$ also a weak conventional superconductor.

\begin{table}[ptb]
\caption{The superconductive parameters of
$N$(\emph{E}$_{\mathrm{{F}}}$) (in unit of states/spin/Ry/cell),
$\omega$$\rm{_{log}}$ (in K), $\uplambda$, and $T_{c}$ (in K) for
some metal-intercalated transition metal dichalcogenides.}
\begin{ruledtabular}
\begin{tabular}{lccccccccccccccccccc}
Compounds&$N$(\emph{E}$_{\mathrm{{F}}}$)&$\omega$$\rm{_{log}}$&$\uplambda$&$T_{c}$&Refs.\\
\hline
TSPTS&23.58&158.0&0.48&1.66&This work\\
PTS&15.14&117.5&0.68&3.84&This work\\
PTSP&20.93&88.3&0.74&3.53&This work\\
PbTiSe$_{2}$&15.72&129.6&0.57&2.44&This work\\
PbTaSe$_{2}$&22.30&99.4&0.66&3.1&\cite{chang2016topological}\\
PbNbSe$_{2}$&25.84&103.4&0.73&4.0&\cite{chen2016ab}\\
SnTaSe$_{2}$&19.86&87.1&0.96&5.7&\cite{chen2016ab}\\
SnNbSe$_{2}$&21.49&72.6&1.28&7.0&\cite{chen2016ab}\\
Na-NbSe$_{2}$&&175.5&0.548&2.993&\cite{lian2017first}\\
\end{tabular}
\end{ruledtabular}
\label{tbl:example}
\end{table}
\par Using a typical value of the effective screened
Coulomb repulsion constant $\mu^{*}$=0.1 as well as our calculated Eliashberg spectral function
$\alpha^{2}$F($\omega$) and $\uplambda$, we can calculate the
logarithmic average frequency $\omega\rm{_{log}}$ and the superconducting
transition temperature $T_{c}$ by
\begin{align}
\omega\rm{_{log}}=exp\left[  \frac{2}{\uplambda}\int_{0}^{\infty}%
\frac{\emph{d}\omega}{\omega}\alpha^{2}\emph{F}(\omega)\rm{log}\omega\right]
\end{align}
and
\begin{align}
T_{c}=\frac{\omega\mathrm{{_{log}}}}{1.2}\mathrm{{exp}\left[  -\frac
{1.04(1+\uplambda)}{\uplambda-\mu^{*}(1+0.62\uplambda)}\right].}%
\end{align}
In Table I, we list the superconductive
parameters of $N$(\emph{E}$_{\mathrm{{F}}}$),
$\omega$$\rm{_{log}}$, $\uplambda$, and $T_{c}$ for TSPTS, PTS, PTSP, and bulk PbTiSe$_{2}$, together with some other metal-intercalated transition metal dichalcogenides \cite{chang2016topological,chen2016ab,lian2017first}. We find that the $T_{c}$ of PbTiSe$_{2}$ is in range of 1.6-3.8 K. Along with increasing the content of Pb, the $\uplambda$ of the 2D film increases, the $\omega\rm{_{log}}$ decreases, and the $T_{c}$ increases rapidly from $\sim$1.6 K to $\sim$3.5 K. The $T_{c}$ of 2.44 K for 3D PbTiSe$_{2}$ is smaller than the maximum $T_{c}$ = 4.5 K for Cu-intercalated TiSe$_{2}$ \cite{morosan2006superconductivity}, but larger than the maximum $T_{c}$ = 1.8 K for pristine TiSe$_{2}$ under pressure \cite{kusmartseva2009pressure}. It is also comparable to that of the newly-found topological superconductor PbTaSe$_{2}$ \cite{chang2016topological}. Our results clearly indicate that the Pb, in addition to Cu, can suppress the CDW and induce superconductivity in TiSe$_{2}$.

\begin{figure}[ptb]
	\centering
	\includegraphics[width=1\linewidth]{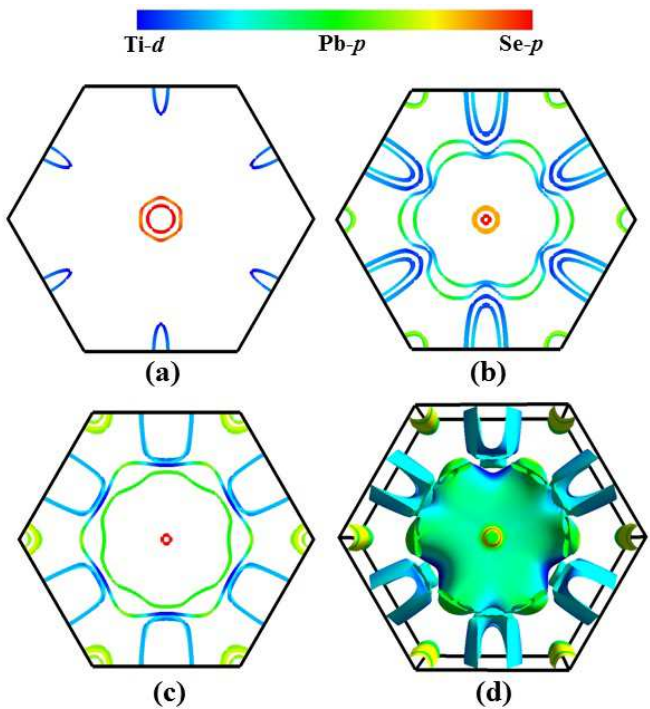}
	\caption{The Fermi surfaces of (a) monolayer 1$T$-TiSe$_{2}$, (b) TSPTS, (c) PTSP, and (d) bulk PbTiSe$_{2}$ calculated with SOC. Ti $d$, Pb $p$, and Se $p$ orbital characters are indicated by blue, green and red colors, respectively. The Fermi surface for bulk PbTiSe$_{2}$ is viewed along
the $\Gamma$-A high symmetry direction. }
	\label{fgr:fig7}
\end{figure}

As indicated by the results of the ARPES \cite{zhao2007evolution}, only the Ti 3\emph{d} band is responsible for the conventional \emph{s}-wave superconductivity in 1\emph{T}-Cu$_{x}$TiSe$_{2}$ \cite{li2007single}. The Ti 3\emph{d} band also plays crucial role in pressure-induced superconductivity in pristine 1\emph{T}-TiSe$_{2}$ \cite{kusmartseva2009pressure,calandra2011charge}. However, in our study of Pb-TiSe$_{2}$, although the Ti 3\emph{d} orbitals behave similar with that in 1\emph{T}-Cu$_{x}$TiSe$_{2}$, the so-called Fermi patches near the $\overline{\rm{M}}$ point expanding with increased Cu/Pb doping (see Fig. 7), the Ti contribution to the EPC is ignorable (Figs. 5 and 6). Besides, as shown in Fig. 7, the Se-4\emph{p} orbitals shrink obviously along with increasing of the Pb content while the Pb-6$p$ orbitals become dominant, especially near the $\overline{\rm{K}}$ point and the half center. Therefore, we can conclude that it is the Pb atom that suppresses the CDW and also inspires the superconductivity in 1\emph{T}-TiSe$_{2}$. The Pb atom intercalates the EPC in this appealing TMD.

\section{CONCLUSIONS}
In summary, we have investigated the structure, electronic
structure, phonon spectrum, EPC, and superconducting properties of
the Pb-intercalated 1\emph{T}-TiSe$_{2}$ using
first-principles calculations. We show that Pb intercalation can suppress the CDW instability of the bulk and monolayer 1\emph{T}-TiSe$_{2}$. We further predict the superconductivity in its 3D and 2D forms which is mainly contributed by the low-frequency phonons associated with the Pb and Se vibrations. The Pb 6\emph{p} orbitals are found crucial for the superconductivity. Upon increasing of the Pb content, the $\uplambda$ and the $T_{c}$ of the 2D form increases while the $\omega\rm{_{log}}$ decreases. Our present findings provide a new choice and also a new mechanism
in realizing superconductor in 1\emph{T}-TiSe$_{2}$ and will inspire further
efforts in modificatory metallic TMD materials.

\begin{acknowledgments}
We acknowledge financial support
from National Natural Science Foundation of China under Grants No.
51371195, No. 11675255, and No. 11634008. The calculations were performed at Supercomputer Centre
in China Spallation Neutron Source.
\end{acknowledgments}

\bibliographystyle{apsrev4-1}
\bibliography{bibl}

\end{document}